# Constant-intensity waves and their modulation instability in non-Hermitian potentials

K. G. Makris[1,2,*], Z. H. Musslimani[3], D. N. Christodoulides[4], and S. Rotter[1]

[1]Institute for Theoretical Physics, Vienna University of Technology, Vienna, Austria, A-1040, EU

[2]Department of Electrical Engineering, Princeton University, Princeton, New Jersey, 08544 USA

[3]Mathematics Department, Florida State University, Tallahassee, Florida, USA

[4]College of Optics–CREOL, University of Central Florida, Orlando, Florida 32816, USA

**In all of the diverse areas of science where waves play an important role, one of the most fundamental solutions of the corresponding wave equation is a stationary wave with constant intensity. The most familiar example is that of a plane wave propagating in free space. In the presence of any Hermitian potential, a wave's constant intensity is, however, immediately destroyed due to scattering. Here we show that this fundamental restriction is conveniently lifted when working with non-Hermitian potentials. In particular, we present a whole new class of waves that have constant intensity in the presence of linear as well as of nonlinear inhomogeneous media with gain and loss. These solutions allow us to study, for the first time, the fundamental phenomenon of modulation instability in an inhomogeneous environment. Our results pose a new challenge for the experiments on non-Hermitian scattering that have recently been put forward.**

Our intuition tells us that stationary waves, which have a constant intensity throughout an extended region of space, can only exist when no obstacles hamper the wave's free propagation. Such an obstacle could be an electrostatic potential for an electronic matter wave, the non-uniform distribution of a dielectric medium for an electromagnetic wave or a wall that reflects an acoustic pressure wave. All of these cases lead to scattering, diffraction and wave interference, resulting in the highly complex variation of a wave's spatial profile that continues to fascinate us in all its different manifestations. Suppressing or merely controlling these effects, which are at the heart of wave physics, is a challenging task, as the quest for a cloaking device [1] or the research in adaptive optics [2], and in wavefront shaping through complex media [3] make us very much aware. New strategies in this direction are thus in high demand and would fall on a fertile ground in many of the different disciplines of science and technology in which wave propagation is a key element.

A new avenue to explore various wave phenomena has recently been opened up when it was realized that waves give rise to very unconventional features when being subject to a suitably chosen spatial distribution of both gain and loss. Such non-Hermitian potential regions [4,5], which serve as sources and sinks for waves, respectively, can give rise to novel wave effects that are impossible to realize with conventional, Hermitian potentials. Examples of this kind, that were meanwhile also realized in the experiment [6-10], are the unidirectional invisibility of a gain-loss potential [11], devices that can simultaneously act as laser and as a perfect absorber [12-14] and resonant structures with unusual features like non-reciprocal light transmission [10] or loss-induced lasing [15-17]. In particular, systems with a so-called parity-time ($\mathcal{PT}$) symmetry [18], where gain and loss are carefully balanced, have recently attracted enormous interest in the context of non-Hermitian photonics [19-23].

Inspired by these recent advances, we show here that for a general class of potentials with gain and loss, it is possible to construct constant-intensity wave solutions. Quite surprisingly, these are solutions to both the paraxial equation of diffraction and the nonlinear Schrödinger equation. In the linear regime, such constant- intensity waves resemble Bessel beams of free space [24]. They carry infinite energy, but retain many of their exciting properties when being truncated by a finite-size input aperture. In the nonlinear regime, this new class of waves turns out to be of fundamental importance, as they provide the first instance to investigate the best known symmetry breaking instability, i.e., the so-called modulational instability (MI) [25-30], in inhomogeneous potentials. Using these new solutions for studying the phenomenon of MI, we find that in the defocusing case, unstable finite size and periodic modes appear causing the wave to disintegrate and to generate a train of complex solitons. In the self-focusing regime, the uniform intensity solution is modulationally unstable for all wavenumbers.

## Results

Our starting point is the well known nonlinear Schrödinger equation (NLSE). This scalar wave equation encompasses many aspects of optical wave propagation as well as the physics of matter waves. Specifically, we will consider the NLSE with a general, non-Hermitian potential $V(x)$ and a Kerr nonlinearity,

$$i\frac{\partial \psi}{\partial z} + \frac{\partial^2 \psi}{\partial x^2} + V(x)\psi + g|\psi|^2 \psi = 0 \quad (1)$$

The scalar, complex valued function $\psi(x,z)$ describes the electric field envelope along a scaled propagation distance $z$ or the wave function of a matter wave as it evolves in time. The nonlinearity can either be self-focusing or defocusing, depending on the sign of $g$. For this general setting, we now investigate a whole family of recently introduced potentials $V(x)$ [31], which are determined by the following relation,

$$V(x) = W^2(x) - i\frac{dW(x)}{dx} \quad (2)$$

where $W(x)$ is a given real function. In the special case where $W(x)$ is even, the actual optical potential $V(x)$ turns out to be $\mathcal{PT}$-symmetric, since $V(x) = V^*(-x)$. We emphasize, however, that our analysis is also valid for confined, periodic or disordered potentials $W(x)$, which do not necessarily lead to a $\mathcal{PT}$-symmetric form of $V(x)$ (but for which gain and loss are always balanced since $\int_{-\infty}^{+\infty} \text{Im}[V(x)]dx = 0$). For the entire non-Hermitian family of potentials that are determined by equation (2) (see Methods) we can prove, that the following analytical and stationary constant-intensity wave is a solution to the NLSE in equation (1),

$$\psi(x,z) = A\, e^{igA^2 z + i\int W(x)dx} \quad (3)$$

notably with a constant and real amplitude A. We emphasize here the remarkable fact that this family of solutions exists in the linear regime ($g=0$) as well as for arbitrary strength of nonlinearity ($g=\pm 1$). Under linear conditions ($g=0$) the constant-intensity wave given by equation (3) is one of the radiation eigenmodes (not confined) of the potential with propagation constant equal to zero. Another interesting point to observe is that the above solutions exist only for non-Hermitian potentials, since for $W(x) \to 0$ we also have $V(x) \to 0$. Therefore, these families of counterintuitive solutions are a direct consequence of the non-Hermitian nature of the involved potential $V(x)$ and as such exist only for these complex structures with gain and loss. The fact that such constant-intensity waves are a direct generalization of the fundamental concept of free-space plane waves to complex environments can be easily understood by setting $W(x) = C = \text{const.}$ with $C \in \mathbb{R}$. In this case the potential $V(x) = C^2$ corresponds to a bulk dielectric medium for which the constant-intensity waves reduce to the plane waves of homogeneous space, $\psi = e^{iCx}$. It can also be shown that the potential $W(x)$ determines the power flow in the transverse plane that physically forces the light to flow from the gain to the loss regions. In particular, the transverse normalized Poynting vector defined as, $S = (i/2)(\psi \, \partial\psi^*/\partial x - \psi^* \, \partial\psi/\partial x)$ takes on the following very simple form: $S = -A^2 W(x)$.

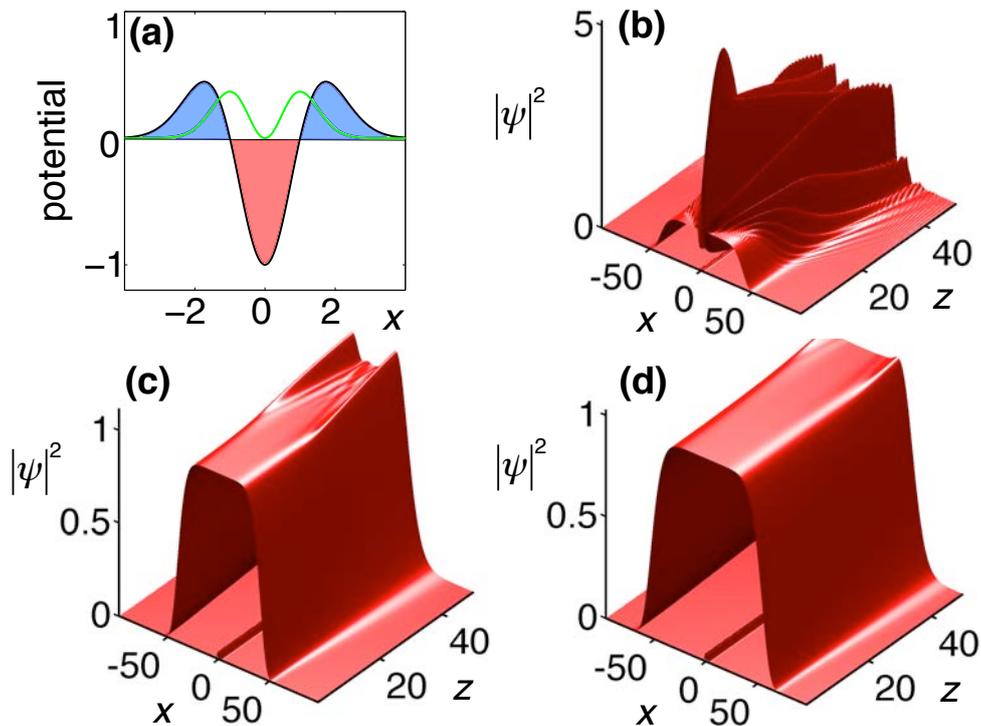

**Figure 1 | Constant-intensity waves in a waveguide coupler.** (**a**) Real part (green line) and imaginary part (black line) of the complex potential $V(x)$ satisfying equation (2) (blue filled regions depict loss whereas the red one depicts gain). (**b**) Evolution of a constant amplitude without the correct phase at the input at $z=0$. (**c,d**) Spatial diffraction of the truncated constant-intensity solution satisfying the correct phase relation of equation (3). Two different input truncations are shown for comparison. The lines in the $x-z$ planes of (**b,c,d**) around $x=0$ depict the real refractive index of the potential as shown in (a).

In order to illustrate the properties of such constant-intensity solutions, we consider the following one-dimensional potentials (not counting the direction of propagation $z$) generated by Hermite polynomials choosing $W(x) = H_n(x)e^{-Bx^2}$. The results for $n=1, B=0.5$ are shown in Fig. 1. Note that the corresponding localized optical potential $V(x)$ is not $\mathcal{PT}$-symmetric (Fig. 1a) and physically describes a waveguide coupler with optical gain in the middle and lossy arms in the evanescent region around it. If the initial beam is not designed to have the correct phase (as given by equation (3)) but is instead $\psi(x,0) = A$ then the light diffracts fast to the gain region, as can be seen in Fig. 1b. In Figs. 1c,d, we show the results for the constant-intensity solutions with the correct phase, where diffraction is found to be strongly suppressed. Similar to the diffraction-free beams [24], we find that the wider the width of the truncation aperture is at the input facet, $z=0$, the larger is the propagation distance after which the beam starts to diffract (compare Fig. 1c with Fig. 1d).

Similar constant-intensity solutions can also be derived in two spatial dimensions $x, y$. The family of these complex potentials $V(x,y)$ and the corresponding constant-intensity solutions $\psi(x,y,z)$ of the two-dimensional NLSE

$$i\frac{\partial \psi}{\partial z} + \frac{\partial^2 \psi}{\partial x^2} + \frac{\partial^2 \psi}{\partial x^2} + V(x,y)\psi + g|\psi|^2 \psi = 0 \text{ are:}$$

$$V(x,y) = |\vec{W}|^2 - i\vec{\nabla} \cdot \vec{W} \quad (4)$$

$$\vec{\nabla} \times \vec{W} = 0 \quad (5)$$

$$\psi(x,y,z) = Ae^{igA^2 z + i\int_C \vec{W} \cdot d\vec{x}} \quad (6)$$

where $\vec{W} \equiv \vec{x} W_x + \vec{y} W_y$ with $W_x, W_y$ being real functions of $x, y$ and $C$ being any smooth open curve connecting an arbitrary point $(a,b)$ to any different point $(x,y)$. As in the one-dimensional case, these solutions are valid in both the linear and the nonlinear domain. For the particular case of $W_x = \cos x \sin y$, $W_y = \cos y \sin x$, the resulting irrotational periodic potential $V(x,y)$ is that of an optical lattice with alternating gain and loss waveguides. The imaginary part of such a lattice is shown in Fig. 2a. In Fig. 2b we display the diffraction of a constant-intensity beam with the correct phase (as in equation 6) launched onto such a lattice through a circular aperture. As we can see, the beam maintains its constant intensity over a remarkably long distance. In Fig. 2c we present the corresponding transverse Poynting vector defined as $\vec{S} = (i/2)(\psi \vec{\nabla}\psi^* - \psi^* \vec{\nabla}\psi)$, illustrating that the wave flux follows stream line patterns from the gain to the loss regions. Once a finite beam starts to diffract, this balanced flow is disturbed and the waves are concentrated in the gain regions.

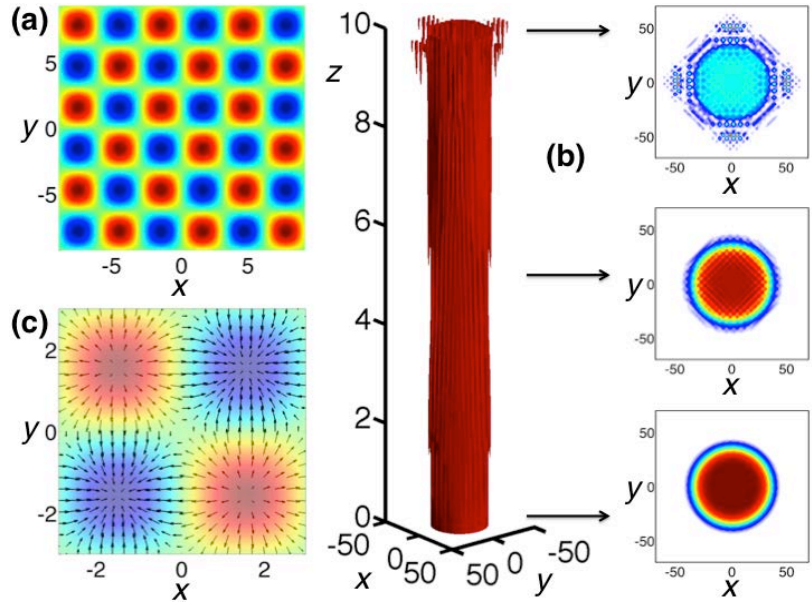

**Figure 2 | Constant-intensity waves in a two-dimensional optical lattice.** (**a**) Imaginary part of the complex potential $V(x,y)$ discussed in the text. Red and blue regions correspond to gain and loss, respectively. (**b**) Iso-contour of the beam intensity launched onto the potential in (a) through a circular aperture of radius ~$40\lambda_0$, where $\lambda_0$ is the free space wavelength. Also shown are three transverse intensity plots (from bottom to top) at $z=0$, $z=5$, $z=10$. (**c**) Transverse power flow pattern (indicated by arrows) of the beam at $z=5$.

Quite remarkably, the above diffraction-free and uniform intensity waves are also solutions of the NLSE for both the self-focusing and the defocusing case. This allows us to study for the first time the modulation instability of such solutions under small perturbations. In particular, we are interested in understanding the linear stability of the solutions of equation (1) of the form $\psi(x,z) = \left[A + \varepsilon F_\lambda(x) e^{i\lambda z} + \varepsilon G_\lambda^*(x) e^{-i\lambda^* z}\right] e^{i\theta(x,z)}$, where the phase function is $\theta(x,z) = gA^2 z + \int W(x)dx$. This expression describes the stationary constant-intensity wave under the perturbation of the eigenfunctions $F_\lambda(x)$ and $G_\lambda(x)$ with $\varepsilon \ll 1$. The imaginary part of $\lambda$ measures the instability growth rate of the perturbation and determines whether a constant-intensity solution is stable ($\lambda \in \mathbb{R}$) or unstable ($\lambda \in \mathbb{C}$). To leading order in $\varepsilon$ we obtain a linear eigenvalue problem for the two-component perturbation eigenmodes $\varphi_\lambda(x) \equiv [F_\lambda(x) \; G_\lambda(x)]^T$ the eigenvalues of which are $\lambda$. This eigenvalue problem and the operator matrix $\vec{M}$ are defined in the Methods section. So far the presented MI-analysis is general and can be applied to any real $W(x)$ (periodic or not). To be more specific, we now apply this analysis to study the MI of constant-intensity waves in $\mathcal{PT}$-symmetric optical lattices [19,20], assuming that $W(x)$ is a periodic potential with period $\alpha$. In particular, we consider the example of a $\mathcal{PT}$-symmetric photonic lattice where $W(x) = \frac{V_0}{2} + V_1 \cos(x)$ (the resulting optical potential and the corresponding constant-intensity solution are given in the Methods section). For all the subsequent results we will always assume (without loss of generality) that $V_0 = 4$ and $V_1 = 0.2$. It is important to note here, that for these parameters our $\mathcal{PT}$-lattice $V(x)$ is in the so-called "unbroken $\mathcal{PT}$-symmetric phase" with only real propagation constants (see Methods). In the broken phase some of these eigenvalues are complex and the instabilities due to nonlinearity are physically expected. Since $W(x)$ is periodic we can expand the perturbation eigenvectors $\varphi_\lambda(x)$ in a Fourier series and construct numerically the bandstructure of the stability problem (different from the physical band-structure of the optical lattice). Based on the above, the Floquet-Bloch theorem implies that the eigenfunctions $\varphi_\lambda(x)$ can be written in the form $\varphi_\lambda(x) = \phi(x,k)e^{ikx}$, where $\phi(x,k) = \phi(x+\alpha,k)$ with $k$ being the Bloch momentum of the stability problem (see Methods). The corresponding results are illustrated in Figs. 3 a,b for a self-focusing nonlinearity ($g=1$) and for different values of the amplitude $A$. More specifically, we show the instability growth rate $|\text{Im}\{\lambda(k)\}|$ as a function of the perturbation eigenvector $k$ in the first half Brillouin zone. We see that the constant-intensity waves are linearly unstable for any value of the Bloch momentum of the imposed perturbation and that instability band gaps form due to the periodic nature of the imposed perturbations. The different bands are illustrated in Figs. 3a,b with different colors.

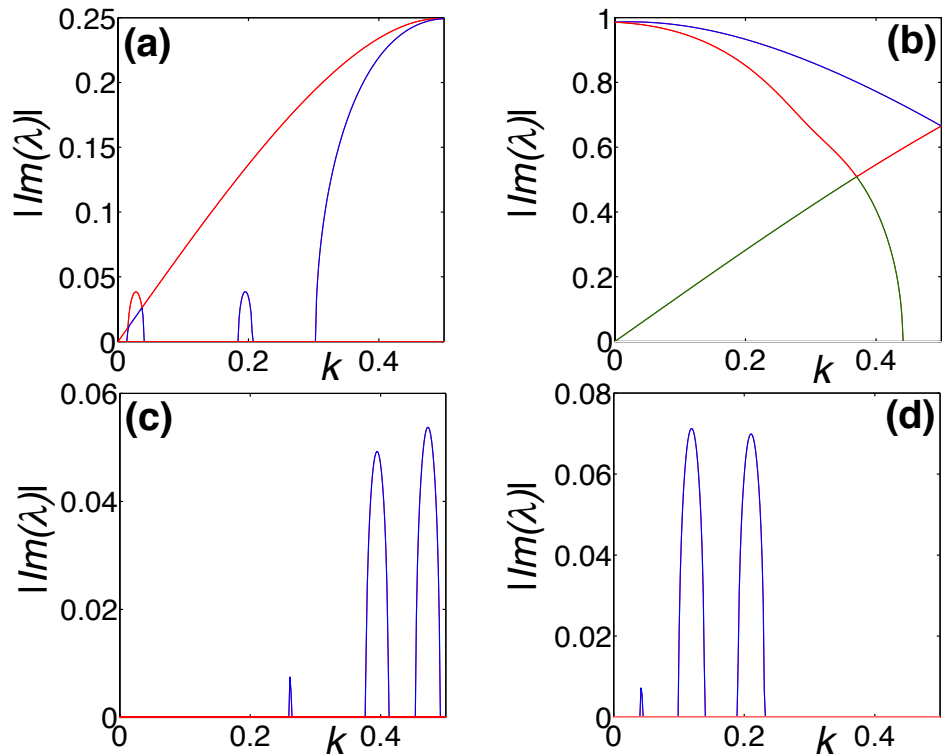

**Figure 3 | Modulation instability diagrams for self-focusing and defocusing nonlinearity.** Growth rate of the instability $|\text{Im}\{\lambda(k)\}|$ as a function of the Bloch momentum (half of first Brillouin zone), for

self-focusing nonlinearity and amplitudes (**a**) $A = 0.5$, (**b**) $A = 1$, and for defocusing nonlinearity (**c**) $A = 1$, and (**d**) $A = 2$. Different colors in (a),(b) denote different instability bands.

The situation is different for the defocusing case ($g = -1$) where the results are presented in Figs. 3c,d. For some values of $k$, the constant-intensity solutions are linearly stable and their instability dependence forms bands reminiscent of the bands appearing in conventional MI results for bulk or periodic potentials [25,28,29], but quite different and profoundly more complex.

In order to understand the physical consequences of such instabilities and how they lead to filament formation, we have performed independent numerical simulations for the dynamics of the constant-intensity solutions against specific perturbations with results being shown in Fig. 4. More specifically, we examine the intensity evolution of a constant-intensity solution when being perturbed by a specific Floquet-Bloch stability mode. In other words, at the input of the waveguide structure at $z = 0$, we have $\psi(x, z = 0) = [A + \varepsilon F_\lambda(x) + \varepsilon G_\lambda^*(x)] e^{i\theta(x,0)}$, with phase $\theta(x) = \frac{V_0 x}{2} + V_1 \sin(x)$ and we are interested whether the linear stability analysis captures the exponential growth of the imposed perturbations correctly. For the considered $\mathcal{PT}$-symmetric lattice with self-focusing nonlinearity we examine the nonlinear dynamics of the constant-intensity solution and the result is presented in Fig. 4a. For a perturbation eigenmode with Bloch momentum $k = 0$ and $A = 1$, $\varepsilon = 0.01$, we can see from Fig. 3b that $\text{Im}\{\lambda(0)\} \sim 1$. Therefore we can estimate the growth for a propagation distance of $z = 6$ to be around $|1 + 0.01 \cdot e^{1 \cdot 5}|^2 \sim 6.1$, which agrees very well with the dynamical simulation of Fig. 4a. Similarly, for the defocusing nonlinearity (Fig. 4b), and for parameters $k = 0.22$ and $A = 2$, $\varepsilon = 0.001$, we estimate the growth for a propagation distance $z = 35$ to be around $\left|2 + 0.001 \cdot e^{0.08 \cdot 35}\right|^2 \sim 4.06$, which matches very well with the numerical propagation result of Fig. 4b.

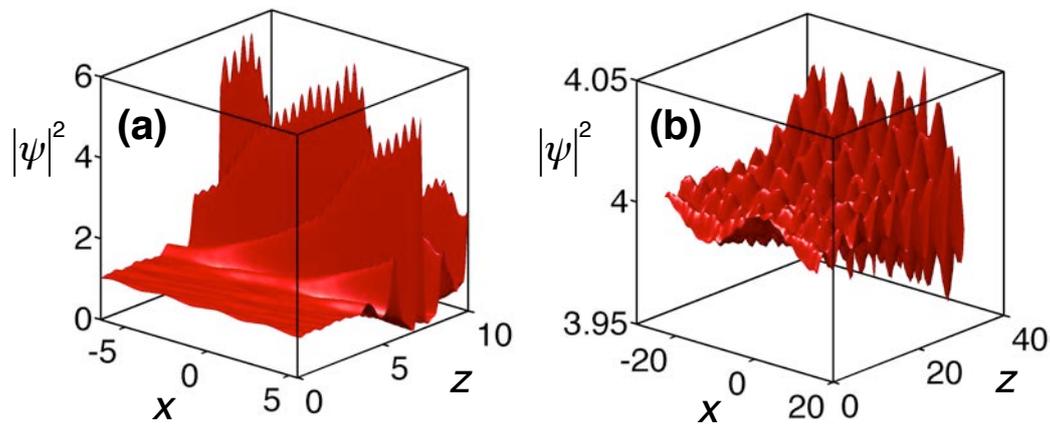

**Figure 4 | Perturbed constant-intensity waves propagating in nonlinear media.** Numerical results for the intensity evolution of a constant-intensity wave for (**a**) a self-focusing nonlinearity with parameters $k = 0$, $A = 1$, $\varepsilon = 0.01$, and (**b**) for a defocusing nonlinearity with parameters $k = 0.22$, $A = 2$, $\varepsilon = 0.001$ The peak values are indicated on the vertical axes and match very well with the results of our perturbation analysis.

## Discussion

Symmetry breaking instabilities belong to the most fundamental concepts of nonlinear sciences. They lead to many rich phenomena such as pattern formation, self-focusing and filamentation just to name a few. The best known symmetry breaking instability is the modulational instability. In its simplest form, it accounts for the break up of a uniform intensity state due to the exponential growth of random perturbations under the combined effect of dispersion/diffraction and nonlinearity. Most of the early work on MI has been related to classical hydrodynamics, plasma physics and nonlinear optics. Soon thereafter, it was realized that the idea of MI is in fact universal and could exist in other physical systems. For example, spatial optics is one particular area that provides a fertile ground where MI can be theoretically modeled (mainly within the framework of the nonlinear Schrödinger equation) and experimentally realized. Indeed, temporal MI has been observed in optical fibers as well as its spatial counterpart in nonlinear Kerr, quadratic, biased photorefractive media with both coherent and partially coherent beams and in discrete waveguide arrays [25-30]. Up to now most of the mathematical modeling of MI processes has been focused on wave propagation in homogeneous nonlinear media where an exact constant-intensity solution for the underlying governing (NLSE type) equations can be obtained. In this context inhomogeneities are considered problematic as they provide severe conceptual limitations that hinder one from constructing a constant-intensity solution, a necessary condition to carry out the MI analysis. Several directions have been proposed to bypass this limitation. They can be organized into three distinct categories: (i) the tight binding approach, in which case the NLSE equation in the presence of an external periodic potential is replaced by its discrete counterpart that, in turn, admits an exact plane wave solution (a discrete Floquet-Bloch mode) (ii) MI of nonlinear Bloch modes and (iii) direct numerical simulations using a broad beam as an initial condition whose nonlinear evolution is monitored. However, none of these alternatives amount to true MI.

We overcome such difficulties by introducing the new family of constant-intensity waves, which exists in a general class of complex optical potentials. These novel type of waves have constant intensity over all space despite the presence of non-Hermitian waveguide structures. They also remain valid for any sign of Kerr nonlinearity and thus allow us, for the first time, to perform a modulational stability analysis for non-homogeneous potentials. The most appropriate context to study the MI of such solutions is that of $\mathcal{PT}$-symmetric optics [6-11,14,19-22]. We find that in the self-focusing regime, the waves are always unstable, while in the defocusing regime the instability appears for specific values of Bloch momenta. In both regimes (self-focusing, defocusing), the constant-intensity solutions break up into filaments following a complex nonlinear evolution pattern.

We expect that our predictions can be verified by combining recent advances in shaping complex wave fronts [3] with new techniques to fabricate non-Hermitian scattering structures with gain and loss [7-10]. Since the precise combination of gain and loss in the same device is challenging, we suggest using passive structures with only loss in the first place. For such suitably designed passive systems [6] solutions exist that feature a pure exponential decay in the presence of an inhomogeneous index distribution. This exponential tail should be observable in the transmission intensity as measured at the output facet of the system. Another possible direction is that of considering evanescently coupled waveguide systems. Using coupled mode theory one can analytically show, that our constant-intensity waves exist also in such discrete systems with distributed gain and loss all over the waveguide channels. In this case the constant-intensity waves are not radiation modes but rather supermodes of the coupled system. With these simplifications an experimental demonstration of our proposal should certainly be within reach of current technology.

# Methods

**Constant-Intensity solutions of the non-Hermitian NLSE**

We prove here analytically that stationary constant-intensity solutions of the NLSE exist for a wide class of non-Hermitian optical potentials (which are not necessarily $\mathcal{PT}$-symmetric). We are looking for solutions of the NLSE of the form $\psi(x,z) = f(x)\exp(i\mu z)$, where $f(x)$ is the complex field profile and $\mu$ the corresponding propagation constant, to be found. By substitution of this last relation into equation (1) we get the following nonlinear equation $-\mu f + f_{xx} + V(x)f + g|f|^2 f = 0$. We assume a solution of the form $f(x) = \rho(x)\exp[i\Theta(x)]$, with $\rho(x), \Theta(x)$ real functions of position $x$. Since $V(x) = V_R(x) + iV_I(x)$, the last nonlinear equation can be separated in real and imaginary parts. As a result we get the following two coupled equations for the real and the imaginary part of the complex potential, respectively:

$$\rho_{xx} + [V_R(x) - \Theta_x^2 - \mu]\rho + g\rho^3 = 0 \quad (7)$$

$$\Theta_{xx}\rho + 2\Theta_x \rho_x + \rho V_I = 0 \quad (8)$$

where $\Theta_x \equiv d\Theta/dx$ and $\rho_x \equiv d\rho/dx$. By choosing $V_R(x) = \Theta_x^2$, and by solving equation (8) to get $V_I(x) = -\Theta_{xx} - 2\Theta_x \dfrac{\rho_x}{\rho}$, we can reduce the above system of coupled nonlinear ordinary differential equations to only one, namely $\rho_{xx} - \mu\rho + g\rho^3 = 0$. If we assume now a constant amplitude solution, namely $\rho(x) = A = const.$, we have the following general solution for any real-valued phase function $\Theta(x)$: $\psi(x,z) = A\exp[i\Theta(x) + igA^2 z]$, where $V_R(x) = \Theta_x^2$, and $V_I(x) = -\Theta_{xx}$. By setting $W(x) \equiv \Theta_x(x)$ we can write the optical potential, for which the constant-intensity solution exists, as $V(x) = W^2(x) - i\,dW/dx$, and the constant-intensity solution itself reads as follows $\psi(x,z) = A\exp\left[igA^2 z + i\int W(x)dx\right]$. We can easily see that in the special case where $W(x)$ is even, the actual optical potential $V(x)$ is $\mathcal{PT}$-symmetric.

**Modulation instability analysis in optical potentials**

In order to study the modulation instability of the uniform intensity states for any given $W$, we consider small perturbation of the solutions of the NLSE of the form: $\psi(x,z) = \left[A + \varepsilon F_\lambda(x)e^{i\lambda z} + \varepsilon G_\lambda^*(x)e^{-i\lambda^* z}\right]e^{i\theta(x,z)}$ where $\theta(x,z) = gA^2 z + \int W(x)dx$ and $\varepsilon \ll 1$. Here, $F_\lambda(x)$ and $G_\lambda(x)$ are the perturbation eigenfunctions and the imaginary part of $\lambda$ measures the instability growth rate of the perturbation. By defining the perturbation two-component eigenmode $\varphi_\lambda(x) \equiv [F_\lambda(x)\ G_\lambda(x)]^T$, we obtain the following linear eigenvalue problem (to leading order in $\varepsilon$):

$$\vec{M}(\hat{L}_\pm) \cdot \varphi_\lambda(x) = \lambda\, \varphi_\lambda(x) \qquad (9)$$

where the operator matrix $\vec{M}(\hat{L}_\pm)$ is defined by the following expression

$$\vec{M}(\hat{L}_\pm) = \begin{pmatrix} \hat{L}_+ & gA^2 \\ -gA^2 & -\hat{L}_- \end{pmatrix} \qquad (10)$$

and the related linear operators are defined by the relationships:

$$\hat{L}_\pm = \hat{L}_0 \pm i\hat{L}_1 \qquad (11)$$

$$\hat{L}_0 = gA^2 + \partial^2/\partial x^2 \qquad (12)$$

$$\hat{L}_1 = 2W(x)\partial/\partial x \quad (13)$$

So far the above discussion is general and applies to any (periodic or not) potential $W(x)$ that is real.

**Properties of the $\mathcal{PT}$-symmetric optical lattice**

We choose a specific example of a well-known non-Hermitian potential, i.e., that of a $\mathcal{PT}$-symmetric optical lattice [19,20]. More specifically for the particular $W(x) = \frac{V_0}{2} + V_1 \cos(x)$, we get the corresponding optical potential and constant-intensity wave:

$$V(x) = \left[\frac{V_0^2}{4} + V_1^2 \cos^2 x + V_0 V_1 \cos x\right] + iV_1 \sin x \quad (14)$$

$$\psi(x,z) = A \exp\left[igA^2 z + i\frac{V_0 x}{2} + iV_1 \sin x\right] \quad (15)$$

It is obvious that this potential is $\mathcal{PT}$-symmetric since it satisfies the symmetry relation $V(x) = V^*(-x)$. In order for the constant-intensity solution to be periodic in $x$ with the same period as the lattice, the constant term $V_0$ must be quantized, namely $V_0 = 0, \pm 2, \pm 4, \ldots$ Another important point is the constant term $V_0$ contained in equation (14). This constant term that appears in the potential $W(x)$ results in another constant term in the actual potential $V(x)$ and can be removed (with respect to the NLSE) with a gauge transformation of the type $\tilde{\psi}(x,z) = \psi(x,z)e^{izV_0^2/4}$. Even though this is the case, this term is important because it also appears in the real part of $V(x)$. It determines if the $\mathcal{PT}$-lattice is in the broken or in the unbroken phase, regarding its eigenspectrum. For the considered parameters, the lattice is below the exceptional point and its eigenvalue spectrum is real.

**Plane Wave Expansion Method**

Even though our methodology is general, we apply it to study the modulation instability of constant-intensity waves in $\mathcal{PT}$-symmetric optical lattices. In particular we consider the periodic $W(x)$ (with period $\alpha$) that leads to equations (14) and (15). Since we are interested in the modulational instabilities of the constant-intensity wave solution of the NLSE under self-focusing and defocusing nonlinearities, we want the $\mathcal{PT}$-lattice $V(x)$ to be in the unbroken phase. In the broken phase some eigenvalues are complex and the instabilities are physically expected. That is the reason why we choose (without loss of generality) the parameters $V_0 = 4$ and $V_1 = 0.2$, which lead to an "unbroken" spectrum with real eigenvalues. Since $W(x)$ is periodic we can expand the perturbation eigenvectors $\varphi_\lambda(x)$ in Fourier series and construct numerically the band-structure of the stability problem. So at this point we have to distinguish between the physical band-structure of the problem and the perturbation band-structure of the stability problem of equation (10). Based on the above, the Floquet-Bloch theorem implies that the eigenfunctions $\varphi_\lambda(x)$ can be written in the form $\varphi_\lambda(x) = \phi(x,k)e^{ikx}$, where $\phi(x,k) = \phi(x+\alpha,k)$ with $k$ being the Bloch momentum of the stability problem. Applying the plane wave expansion method, the wavefunctions $\phi(x,k)$ and the potential $W(x)$ can be expanded in Fourier series as:

$$\varphi_\lambda(x) = \sum_{n=-\infty}^{+\infty} \begin{pmatrix} u_n(k) \\ v_n(k) \end{pmatrix} e^{i(qn+k)x} \quad (16)$$

$$W(x) = \sum_{n=-\infty}^{+\infty} W_n e^{iqnx} \quad (17)$$

where $q = 2\pi/\alpha$ is the dual lattice spacing. Substitution of equation (16) and equation (17) into the eigenvalue problem of equation (10), leads us to the following nonlocal system of coupled linear eigenvalue equations for the perturbation modes $u_n, v_n$ and the band eigenvalue $\lambda(k)$ that depends on the Bloch momentum $k$:

$$\Omega_n(k)u_n - \sum_{n=-\infty}^{+\infty} U_{n,m}(k)u_{n-m} + gA^2 v_n = \lambda u_n$$
$$-gA^2 u_n - \Omega_n(k)v_n - \sum_{m=-\infty}^{+\infty} U_{n,m}(k)v_{n-m} = \lambda v_n \quad (18)$$

where $U_{n,m}(k) = 2[q(n-m)+k]W_m$ and $\Omega_n(k) = gA^2 - (qn+k)^2$. The family of constant-intensity wave solutions of the NLSE is modulationally unstable if there exists a wave number $k$ for which $\mathrm{Im}\{\lambda(k)\} \neq 0$, while they are stable if $\lambda(k)$ is real. For our case the periodic function $W$ is given by $W(x) = \frac{V_0}{2} + V_1 \cos(qx)$ for which equation (18) becomes:

$$\mu_n(k)u_n - a_{n-1}(k)u_{n-1} - a_{n+1}(k)u_{n+1} + gA^2 v_n = \lambda u_n$$
$$-gA^2 u_n - v_n(k)v_n - a_{n-1}(k)v_{n-1} - a_{n+1}(k)v_{n+1} = \lambda v_n \quad (19)$$

where $a_n(k) = V_1(qn+k)$, $\mu_n(k) = \Omega_n(k) - V_0(qn+k)$ and $v_n(k) = \Omega_n(k) + V_0(qn+k)$.

**Direct Eigenvalue Method**

An alternative way (instead of the plane wave expansion method that was used above) of solving the infinite dimensional eigenvalue problem of equation (10) is to directly apply the Floquet-Bloch theorem on the eigenfunctions $\varphi_\lambda(x)$, employ the Born-von-Karman boundary conditions (periodic boundary conditions at the endpoints of the finite lattice) and construct numerically the bandstructure of the instability growth for every value of the Bloch momentum. In particular, the eigenfunctions can be written as $\varphi_\lambda(x) = [u(x)e^{ikx} \quad v(x)e^{ikx}]^T$, where $u(x) = u(x+\alpha), v(x) = v(x+\alpha)$. Substituting this form of the perturbation eigenfunctions into equation (10), we get the following eigenvalue problem:

$$\begin{pmatrix} \hat{L}_+ + 2ik\partial/\partial x - k^2 - 2kW(x) & gA^2 \\ -gA^2 & -\hat{L}_- - 2ik\partial/\partial x + k^2 - 2kW(x) \end{pmatrix} \cdot \begin{pmatrix} u \\ v \end{pmatrix} = \lambda(k) \begin{pmatrix} u \\ v \end{pmatrix} \quad (20)$$

where the Bloch momentum takes values in the first Brillouin zone $k \in [-2\pi/\alpha, 2\pi/\alpha]$ and the operators $\hat{L}_+, \hat{L}_-$ are defined by equations (11)-(13). By applying the finite difference method and spectral differentiation matrices (both methods lead to results in excellent agreement) we restrict our analysis to one unit cell $x \in [-\alpha/2, \alpha/2]$, in order to calculate the growth rate of the random perturbations for every value of the Bloch momentum. We have checked explicitly that both approaches, i.e., the plane wave expansion method based on equation (19) and the direct eigenvalue analysis based on equation (20), give the same results.

**Analytical results in the shallow lattice limit**

In the limit of a shallow optical lattice (the refractive index difference between the periodic modulation and the background refractive index value is very small), one can gain substantial insight into the structure of the unstable band eigenvalues by deriving an approximate analytical expression for $\lambda(k)$ valid near the Bragg points based on degenerate perturbation theory. These points are given by:

$$\lambda_{n+}(k) = \sqrt{(k+nq)^2\left[(k+nq)^2 - k_c\right]} \quad (21)$$

$$\lambda_{n-}(k) = \sqrt{(k-nq)^2\left[(k-nq)^2 - k_c\right]} \quad (22)$$

$$\lambda_0(k) = \sqrt{k^2\left[k^2 - k_c\right]} \quad (23)$$

where $k_c = 2A^2$ and $n = 1,2,3,\ldots$. The above analytical formulas lead to an excellent match with the numerical approaches in the shallow lattice limit ($V_1 \ll 1$).

**Complex filament formation**

In order to understand better the complex filament formation of a constant-intensity solution in a $\mathcal{PT}$-symmetric lattice for both signs of nonlinearity, we performed nonlinear wave propagation simulations based on a spectral fast Fourier approach of the integrating factors method for NLSE. The initial conditions that were used to examine the filament formation were based on the perturbation eigenmode profiles. In particular, we have at $z=0$, the following initial field profile in terms of Bloch eigenfunctions $u(x), v(x)$:

$\psi(x,z=0) = \left[A + \varepsilon u_k(x)e^{ikx} + \varepsilon v_k^*(x)e^{-ikx}\right]e^{i\theta(x)}$, for specific values of Bloch momentum $k$ and the constant-intensity wave amplitude $A$.

## Acknowledgments

K.G.M. is supported by the People Programme (Marie Curie Actions) of the European Union's Seventh Framework Programme (FP7/2007-2013) under REA grant agreement number PIOF-GA-2011- 303228 (project NOLACOME). Z.H.M. was supported in part by NSF Grant No. DMS-0908599. The work of D.N.C. was partially supported by NSF Grant No. ECCS-1128520 and the Air Force Office of Scientific Research Grants Nos. FA9550-12-1-0148 and FA9550-14-1-0037. S.R. acknowledges financial support by the Austrian Science Fund (FWF) through Project SFB NextLite (F49-P10) and Project GePartWave (I1142).

## Author Contributions

All authors have contributed to the development and/or implementation of the concept, discussed and analyzed the results. K.G.M. and Z.M. carried out the analytical and numerical calculations. S.R. and D.C. provided theoretical and conceptual support. K.G.M., Z.M. and S.R. wrote the manuscript with input from all authors.

## Additional Information

No Supplementary Information accompanies the paper.

Competing financial interests: The authors declare no competing financial interests.